\documentclass[final,1p,dvipdfmx]{elsarticle}

\usepackage{amssymb}
\usepackage{color}

\usepackage{lineno}
\usepackage{ulem}

\journal{Physics Letters B}

\begin{document}

\begin{frontmatter}



\title{Non-strange dibaryons studied in the $\gamma{d}${$\to$}$\pi^0\pi^0{d}$ reaction}


\author[elph]{T.~Ishikawa\corref{cor}}
\ead{ishikawa@lns.tohoku.ac.jp}
\author[elph]{H.~Fujimura\fnref{hf}}
\author[elph]{H.~Fukasawa}
\author[elph]{R.~Hashimoto\fnref{rh}}
\author[elph]{Q.~He\fnref{qh}}
\author[elph]{Y.~Honda}
\author[yama]{T.~Iwata}
\author[elph]{S.~Kaida}
\author[aoba]{H.~Kanda\fnref{hk}}
\author[elph]{J.~Kasagi}
\author[gaku]{A.~Kawano}
\author[elph]{S.~Kuwasaki}
\author[aoba]{K.~Maeda}
\author[tokyo]{S.~Masumoto}
\author[elph]{M.~Miyabe}
\author[elph]{F.~Miyahara\fnref{fm}}
\author[elph]{K.~Mochizuki}
\author[elph]{N.~Muramatsu}
\author[elph]{A.~Nakamura}
\author[elph]{K.~Nawa}
\author[elph]{S.~Ogushi}
\author[elph]{Y.~Okada}
\author[elph]{K.~Okamura}
\author[elph]{Y.~Onodera}
\author[kek]{K.~Ozawa}
\author[gaku]{Y.~Sakamoto}
\author[elph]{M.~Sato}
\author[elph]{H.~Shimizu}
\author[elph]{H.~Sugai\fnref{hs}}
\author[elph]{K.~Suzuki\fnref{ks}}
\author[yama]{Y.~Tajima}
\author[elph]{S.~Takahashi}
\author[elph]{Y.~Taniguchi}
\author[elph]{Y.~Tsuchikawa\fnref{yt}}
\author[elph]{H.~Yamazaki\fnref{hy}}
\author[elph]{R.~Yamazaki}
\author[yama]{H.Y.~Yoshida}
\address[elph]{Research Center for Electron Photon Science (ELPH), Tohoku University, Sendai 982-0826, Japan}
\address[yama]{Department of Physics, Yamagata University, Yamagata 990-8560, Japan}
\address[aoba]{Department of Physics, Tohoku University, Sendai 980-8578, Japan}
\address[gaku]{Department of Information Science, Tohoku Gakuin University, Sendai 981-3193, Japan}
\address[tokyo]{Department of Physics, University of Tokyo, Tokyo 113-0033, Japan}
\address[kek]{Institute of Particle and Nuclear Studies, High Energy Accelerator Research Organization (KEK), Tsukuba 305-0801, Japan}

\cortext[cor]{Corresponding author. Tel.: +81 22 743 3400; fax: +81 22 743 3401.}
\fntext[hf]{Present address: Department of Physics, Wakayama Medical University, Wakayama 641-8509, Japan}
\fntext[rh]{Present address: Institute of Materials Structure Science (IMSS), High Energy Accelerator Research Organization (KEK), Tsukuba 305-0801, Japan}
\fntext[qh]{Present address: Institute of Fluid Physics, China Academy of Engineering (CAEP), Mianyang 621900, China}
\fntext[hk]{Research Center for Nuclear Physics (RCNP), Osaka University, Ibaraki 567-0047, Japan}
\fntext[fm]{Present address: Accelerator Laboratory, High Energy Accelerator Research Organization (KEK), Tsukuba 305-0801, Japan}
\fntext[hs]{Present address: Gunma University Initiative for Advanced Research (GIAR), Maebashi 371-8511, Japan}
\fntext[ks]{Present address: The Wakasa Wan Energy Research Center, Tsuruga 914-0192, Japan}
\fntext[yt]{Present address: Department of Physics, Nagoya University, Nagoya 464-8602, Japan}
\fntext[hy]{Present address: Radiation Science Center, High Energy Accelerator Research Organization (KEK), Tokai 319-1195, Japan}
\begin{abstract}
Coherent double neutral-pion photoproduction on the deuteron, 
$\gamma{d}${$\to$}$\pi^0\pi^0{d}$, has been experimentally 
studied at incident photon energies ranging from 0.75 to 1.15 GeV.
The total cross section as a function of the $\gamma{d}$ center-of-mass
energy shows resonance-like behavior, which peaks at approximately
2.47 and 2.63 GeV.
The measured angular distribution of deuteron emission is
rather flat, which cannot be reproduced by the
kinematics of quasi-free $\pi^0\pi^0$ production with deuteron coalescence.
In $\pi^0d $ invariant-mass distributions,
a clear peak is observed at 
$2.14{\pm}0.01$ GeV$/c^2$ 
with a width of $0.09{\pm}0.01$ GeV$/c^2$.
The spin-parity of this state is restricted to $1^+$, $2^+$ or $3^-$ from 
the angular distributions of the two $\pi^0$s.
The present work shows strong evidence for the existence
of an isovector dibaryon resonance
with a mass of 2.14 GeV$/c^2$.
The $2^+$ assignment is consistent 
with the theoretically predicted ${\cal{D}}_{12}$ 
state, and also with
the energy dependence of the $\pi{d}$ partial-wave 
amplitude $^3\!P_2$ for the $\pi^{\pm}d${$\to$}$\pi^{\pm}d$ and $\pi^+d${$\to$}${pp}$ reactions.
\end{abstract}


\end{frontmatter}

The study of two-baryon systems (dibaryons) has a long history~\cite{cle}.
Since the quark picture of a nucleon was established, a dibaryon has become
a more interesting object to investigate a phase change of 
its basic configuration 
from a molecule-like state consisting of two baryons 
(such as the deuteron) to a hexaquark hadron state, 
which is expected to appear as a spatially-compact
exotic particle.
Understanding dibaryons would not only give a clue to the solution
of the current problem in hadron physics, but also provide an insight into the nuclear equation of state and the interior of a neutron star~\cite{nstar}.

Recently, a dibaryon resonance $d^*(2380)$ 
with $M${$=$}$2.37$ GeV$/c^2$, $\Gamma${$=$}$0.07$ GeV$/c^2$,
and $I\left(J^\pi\right)${$=$}$0\left(3^+\right)$ 
has been
observed in the $pn${$\to$}$\pi^0\pi^0{d}$ reaction by the CELSIUS/WASA 
and WASA-at-COSY 
collaborations~\cite{cels,cosy1}.
The $d^*(2380)$ may be attributed  to an isoscalar $\Delta\Delta$ 
quasi-bound state ${\cal{D}}_{03}$, which was predicted  
by Dyson and Xuong~\cite{dx}
as a member of the sextet non-strange dibaryons 
 ${\cal{D}}_{IJ}$ with isospin $I$ and spin $J$:
${\cal{D}}_{01}$,
${\cal{D}}_{10}$,
${\cal{D}}_{12}$,
${\cal{D}}_{21}$,
${\cal{D}}_{03}$, and
${\cal{D}}_{30}$.
The total cross sections were measured for the $\gamma{d}${$\to$}$\pi^0\pi^0{d}$ reaction
below the incident energy of 0.88 GeV 
at the Research Center for Electron Photon Science (ELPH), 
Tohoku University~\cite{dpipi-plb}.
A slight enhancement corresponding to $d^*(2380)$ was observed 
in the excitation function for 
the $\gamma{d}$ center-of-mass (CM) energy
$W_{\gamma{d}}${$=$}$2.38$--2.61 GeV although it was not statistically 
significant.
A preliminary result obtained by the A2 collaboration 
at the Mainz MAMI accelerator also 
apparently showed a similar enhancement
above an incident energy of 0.41 GeV ($W_{\gamma d}=2.22$ GeV)~\cite{guenther}.

It is important to establish the excitation spectrum of dibaryons 
in order to understand their internal structures. 
Many experimental investigations of the sextet members have been made,
and 
candidates for almost all the members seem to be found.
Dyson and Xuong predicted them using the masses of the 
three experimentally observed states:
the deuteron ${\cal{D}}_{01}$, 
the $^1\!S_0$-$NN$ virtual state ${\cal{D}}_{10}$, 
and a resonance-like structure corresponding to ${\cal{D}}_{12}$
which appears at $M${$=$}$2.16$ GeV$/c^2$ in the $\pi^+{d}${$\to$}${pp}$ reaction~\cite{d12b}.
Moreover, in addition to the observation of a ${\cal{D}}_{03}$ candidate $d^*(2380)$,
an enhancement corresponding to an isotensor $\Delta N$ dibaryon, ${\cal{D}}_{21}$,
is observed in the quasi-free $pp\to \pi^+\pi^-pp$ reaction~\cite{d21},
and a hint of an $I=3$ $\Delta\Delta$ dibaryon,
${\cal{D}}_{30}$, is obtained in the 
$pp\to \pi^-\pi^- \pi^+\pi^+ pp$ reaction~\cite{d30}.
However, the dibaryonic interpretations of the experimental data 
for some members are still questionable.

Theoretically, 
Mulders et al. studied the sextet dibaryons using a bag model~\cite{dib1,dib3}.
They predict that ${\cal{D}}_{03}$ has  $M${$=$}$2.36$ GeV$/c^2$ and 
strongly (weakly) couples to the $^7\!S_3$-$\Delta\Delta$  ($^3\!D_3$-$NN$) state.
They also predict that ${\cal{D}}_{12}$ has  $M${$=$}$2.36$ GeV$/c^2$ and
couples to the ${}^1\!D_2$-$NN$ and ${}^5\!S_2$-$N\Delta$ states.
Gal and Garcilazo analyzed
${\cal{D}}_{03}$ and ${\cal{D}}_{12}$
using three-body hadronic models~\cite{dib4}.
They obtain  $M${$=$}${2.38}$ ($2.15$) GeV$/c^2$ 
for ${\cal{D}}_{03}$ (${\cal{D}}_{12}$)  by solving $\pi{N}\Delta$ ($\pi{NN}$)
 Faddeev equations.
Platonova and Kukulin predicts the ${\cal D}_{03} \to \pi {\cal D}_{12}$
decay, and additional isoscalar and isovector dibaryons~\cite{dib5}.
Experimentally,
${\cal{D}}_{12}$ is given as the $^3\!P_2$ multipole strength 
at  $M${$=$}$2.18$ GeV/$c^2$
in $\pi^{\pm}d$ elastic scattering
by a partial-wave analysis~\cite{pid-pid}.
The corresponding  $^1\!D_2$-$pp$ amplitude
also shows the same structure in the $\pi^+{d}${$\to$}${pp}$ 
reaction~\cite{pid-pp}.
The SAID group provides 
a pole for ${\cal D}_{12}$~\cite{pp-pp} from 
a combined analysis of $\pi d$ elastic scattering, the $\pi d\to pp$ reaction,
and $pp$ elastic scattering.
Hoshizaki shows the ${\cal D}_{12}$ state is required to explain the $pp$ and $\pi d$ phase parameters~\cite{add2a},
and excludes the interpretation of the state as a cusp or a virtual state~\cite{add2b}.
Platonova and Kukulin also claim that conventional meson-exchange models 
do not explain the $pp\to \pi^+d$ reaction~\cite{add3}.
Recently, preliminary results for ${\cal{D}}_{12}$ candidates
observed in the $\gamma{d}${$\to$}$\pi^+\pi^-{d}$ reaction
are reported~\cite{jlab,nks2},
showing a peak with $M${$=$}$2.1$--2.2 GeV$/c^2$ and
$\Gamma${$\simeq$}${0.1}$ GeV/$c^2$ in both the $\pi^{\pm}d$ invariant-mass distributions.
The peak position is close to the sum of the $N$ and $\Delta$ masses.

There is no doubt that the $\pi{d}$ system has a resonance-like structure 
around 2.15 GeV$/c^2$.
If the existence of this resonance ${\cal D}_{12}$
 is verified, our understanding of the sextet dibaryons would be strongly deepened.
To study ${\cal D}_{12}$, the 
$\gamma{d}${$\to$}$\pi^0{d}$ reaction is a convenient approach. 
However, the resonance, if observed, can be also understood as a quasi-free (QF) 
$\Delta$ excitation from a nucleon in the deuteron. 
To find ${\cal D}_{12}$
in the $\pi^0d$ system through the $\gamma{d}${$\to$}${\pi^0\pi^0}d$
reaction is more advantageous, because the QF $\Delta$ excitation is kinematically separable: 
the kinetic energy given to the deuteron is
very small in most cases for QF $\pi^0\pi^0$ production on a nucleon followed by 
deuteron coalescence
 (QFC). In addition, a generated 
$\pi^0{d}$ resonance
following $\pi^0$ emission 
requires an isoscalar coupling in
the initial $\gamma{d}$ state. 
This constraint in addition to a very small $\gamma \pi^0$
coupling may reduce contributions from non-resonance processes. 
In this Letter, we study the $\gamma{d}${$\to$}$\pi^0\pi^0d$ reaction, aiming to observe ${\cal D}_{12}$
in the $\pi^0d$ system $(I${$=$}$1)$ through $\pi^0$ decay from a possible higher-mass 
dibaryon in the $\pi^0\pi^0d$ system $(I${$=$}$0)$.

A series of experiments~\cite{exp}
were carried out using a  bremsstrahlung photon beam from 
1.20-GeV circulating electrons 
in a synchrotron~\cite{stb} at ELPH.
The photon beam is provided by
inserting a carbon fiber
into the circulating electrons~\cite{tag2,bpm}.
The energy of each photon is determined by
detecting the post-bremsstrahlung
electron with a photon-tagging counter, STB-Tagger II.
The tagging energy of the photon beam ranges from 0.75 to 1.15 GeV.
The target used in the experiments was liquid deuterium with a thickness of 45.9~mm.
All the final-state particles in the $\gamma{d}${$\to$}$\pi^0\pi^0d$ reaction
were measured with the FOREST detector~\cite{forest}.
FOREST consists of three different electromagnetic calorimeters (EMCs):
192 CsI crystals, 
252 lead scintillating-fiber modules, and
62 lead-glass counters.
A plastic-scintillator hodoscope (PSH) is placed in front of each EMC
to identify charged particles.
FOREST covers the solid angle of $\sim${88\%} in total.
The typical photon-tagging rate was 20~MHz,
and the photon transmittance (the so-called tagging efficiency)
was $\sim${53\%}~\cite{tag2}.
The trigger condition of the data acquisition (DAQ),
which required to detect more than one final-state particles in coincidence with a 
photon-tagging signal~\cite{forest},
was the same as that in Ref.~\cite{dpipi-plb}.
The average trigger rate was 1.7~kHz, 
and the average DAQ efficiency was 79\%.

Event selection is made 
for the $\gamma{d}${$\to$}$\pi^0\pi^0{d}${$\to$}$\gamma\gamma\gamma\gamma{d}$ reaction.
Initially, events containing four neutral particles and a charged particle 
are selected.
The time difference between every two neutral EMC clusters
out of four is required to be less than three times that
of the time resolution for the difference.
The charged particles are detected with the forward PSH.
The time delay from the response of the four neutral clusters 
is required to be longer than 1 ns.
The deposit energy of a charged particle in PSH is
required to be greater than twice 
that of the minimum ionizing particle.
Further selection 
is made by applying a kinematic fit with six constraints:
energy and three-momentum conservation,
and every $\gamma\gamma$ invariant-mass being the $\pi^0$ mass.
The momentum of the charged particle is obtained from the time delay
assuming that the charged particle has the deuteron mass.
Events for which the $\chi^2$ probability is higher
than 0.4 are selected to reduce those
from other background processes.
Events from deuteron misidentification in the most 
competitive QF $\gamma{p'}${$\to$}$\pi^0\pi^0p$  reaction
are less than 3\%.
Finally, sideband-background subtraction is performed
for accidental-coincidence events detected in STB-Tagger II and FOREST.

\begin{figure}[htbp]
\begin{center}
\includegraphics[width=0.95\textwidth]{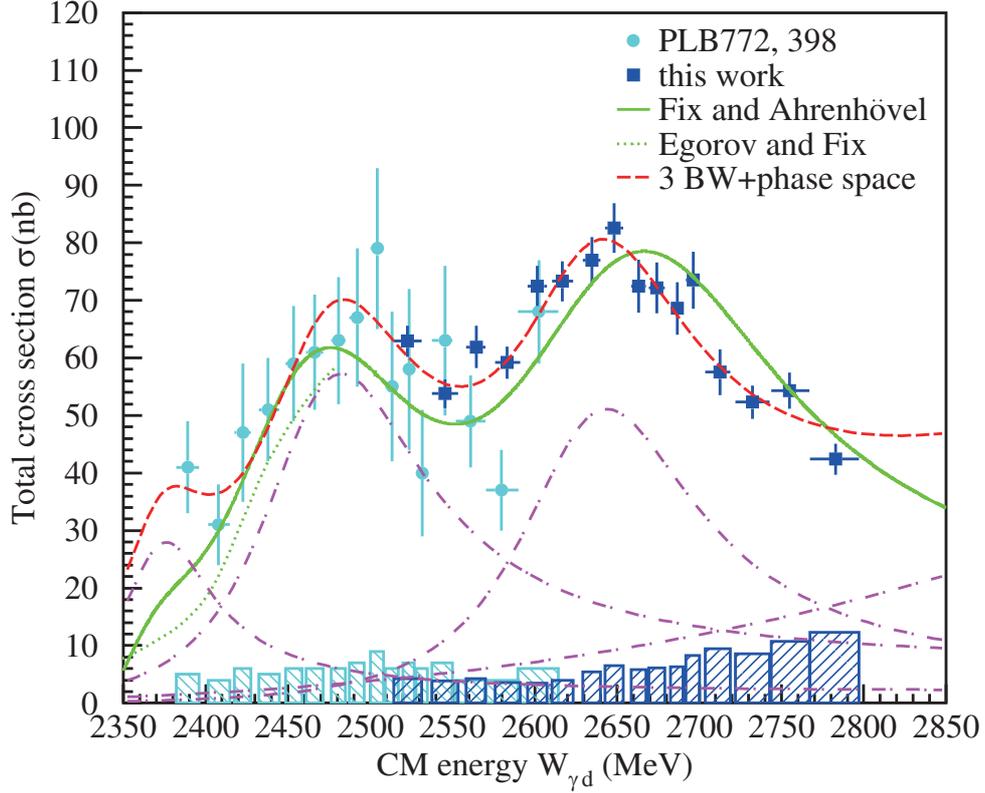}
\end{center}
\caption{Total cross section $\sigma$ as a function of $W_{\gamma{d}}$.
The squares (blue) show $\sigma$ obtained in this work,
while the circles (cyan) show that presented in Ref.~\cite{dpipi-plb}.
The horizontal error of each data point corresponds to 
the coverage of the incident photon energy, and the vertical error shows 
the statistical error of $\sigma$.
The solid and dotted curves (green) show theoretical calculations given in Ref.~\cite{fix1}
and \cite{fix2}, respectively.
The dashed curve (red) shows the fitted function in Eq.~(\ref{eq:3bw}):
a sum of three BW peaks and phase-space contributions.
Each contribution to it is shown in a 
dash-dotted curve (magenta).
The lower hatched histograms (blue and cyan) show the systematic errors of $\sigma$
in this work and in Ref.~\cite{dpipi-plb}, respectively.
}\label{fig1}
\end{figure}

The total cross section is obtained 
by estimating the acceptance of
$\gamma\gamma\gamma\gamma{d}$ detection in 
a Monte-Carlo simulation based on Geant4~\cite{geant4}.
Here,
the event generation is modified from pure phase-space generation
to reproduce the following three measured distributions:
the $\pi\pi$ invariant mass $M_{\pi\pi}$, 
the $\pi{d}$ invariant mass $M_{\pi{d}}$, and 
the deuteron emission angle $\cos\theta_d$ in the $\gamma{d}$-CM
frame.
Fig.~\ref{fig1} shows the total cross section $\sigma$ 
as a function of $W_{\gamma{d}}$.
The data obtained in this work
are consistent with the previously obtained data~\cite{dpipi-plb} within errors.
The systematic uncertainty of $\sigma$ is also given in Fig.~\ref{fig1}. 
It includes the uncertainty of event selection in the kinematic fit, 
that of acceptance owing to the uncertainties of the
$M_{\pi\pi}$, $M_{\pi{d}}$, and $\cos\theta_d$ 
distributions in event generation of the simulation,
that of detection efficiency of deuterons,
and that of normalization resulting from the 
numbers of target deuterons and incident photons.

The excitation function is not monotonically
increasing but shows resonance-like behavior peaked at around
2.47 and 2.63 GeV.
The two-peak structure is similar to the excitation 
function of the $\gamma{N}${$\to$}$\pi^0\pi^0N$ reaction 
with two peaks at 
the $\gamma{N}$-CM energy $W_{\gamma{N}}$
of $\sim${1.5} and $\sim${1.7} GeV~\cite{graal,a2-pi0pi0}, 
corresponding to the second- 
and third-resonance regions of the nucleon. 
A naive interpretation of this behavior may be a QF excitation 
of the nucleon in the deuteron. The solid line in Fig.~\ref{fig1} shows a calculation 
performed by Fix and Ahrenh\"ovel (FA) based on the QFC
mechanism~\cite{fix1}. 
The calculation reproduces the data surprisingly well. 
A calculation performed by Egorov and Fix based~\cite{fix2} on the QFC mechanism
also gives a similar excitation function as shown in Fig.~\ref{fig1} (dotted line).
However, as discussed later, the kinematic condition for the obtained data completely
differs from the QFC process. 

\begin{figure}[htbp]
\begin{center}
\includegraphics[width=0.95\textwidth]{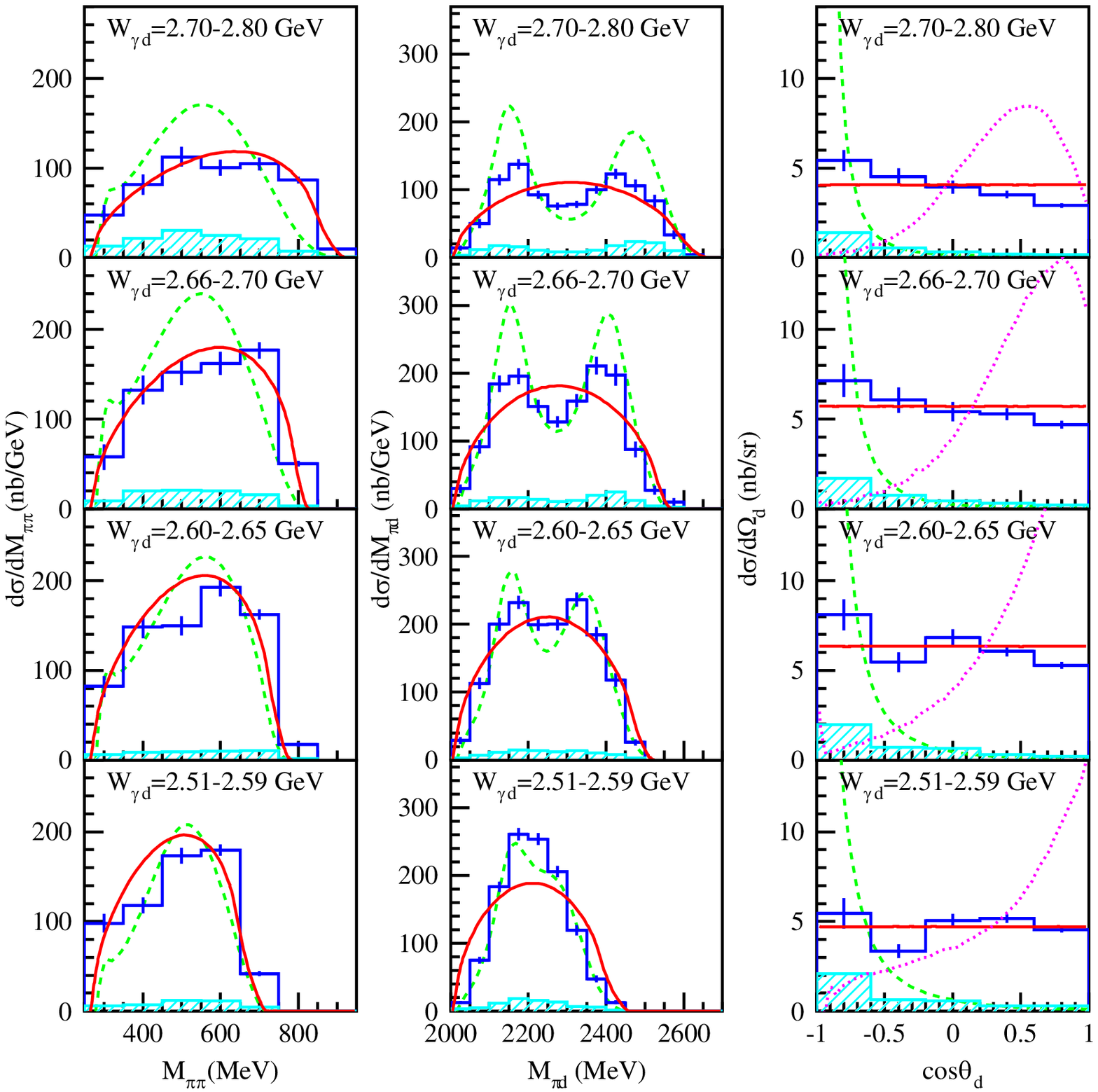}
\end{center}
\caption{
Differential cross sections $d\sigma/dM_{\pi\pi}$ (left), $d\sigma/dM_{\pi{d}}$
(central),
and  $d\sigma/d\Omega_d$ (right) for all the four photon-tagging groups.
The lower hatched histograms (cyan) show 
the corresponding systematic errors.
The dashed curves (green) show the FA calculations based on the QFC mechanism.
The solid (red) and dotted (magenta) curves correspond to the pure phase-space
and semi-QF processes, respectively,
where their yields are normalized so that the total cross section should be the same.
}\label{fig2}
\end{figure}

\begin{figure}[htbp]
\begin{center}
\includegraphics[width=0.9\textwidth]{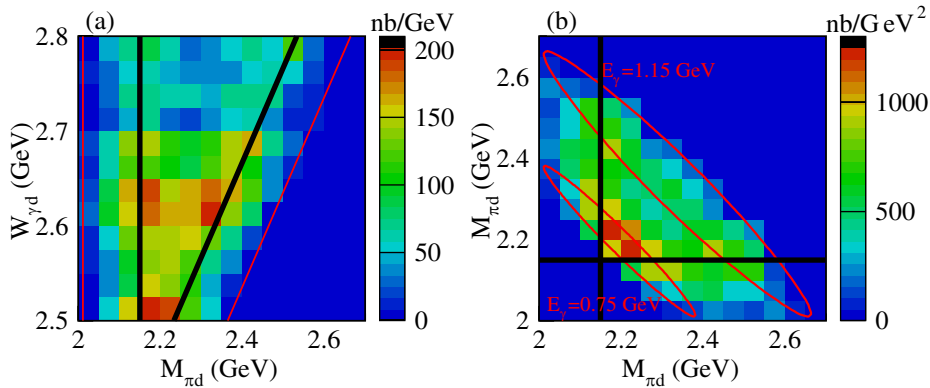}
\end{center}
\caption{
(a) Correlation between $W_{\gamma d}$ and $M_{\pi d}$ 
($d\sigma/dM_{\pi d}$).
The narrow lines (red)
 represent the maximum and minimum values of $M_{\pi d}$ at a fixed 
incident photon energy, and
the bold lines (black) 
show the locus $M_{\pi d}=2.15$ GeV$/c^2$ and its reflection.
(b) Correlation between two $M_{\pi d}$s ($d^2/dM_{\pi_1 d}/dM_{\pi_2 d}$).
The curves (red) represent the boundaries of phase space at the incident energies of 
750 and 1150 MeV.
The bold lines (black) show the loci $M_{\pi d}=2.15$ GeV$/c^2$.
}\label{fig3}
\end{figure}

The differential cross sections, $d\sigma/dM_{\pi\pi}$,
$d\sigma/dM_{\pi{d}}$,
and $d\sigma/d\Omega_d$, are obtained
for each group of photon-tagging channels divided into four groups
as shown in Fig.~\ref{fig2}.
The experimental data are presented by histograms with statistical errors,
the systematic uncertainties by hatched histograms and 
FA calculations by the solid curves.
The $d\sigma/dM_{\pi\pi}$ shows no prominent feature, 
increasing monotonically with increase of $M_{\pi\pi}$ from the 
minimum to the maximum of the available energy.
In contrast, the momenta of the two $\pi^0$s are correlated in the QFC mechanism.
This is partly because the quasi-free $\gamma n\to \pi^0\pi^0n$ reaction 
shows an enhancement in the central part of the $M_{\pi\pi}$ spectrum~\cite{graal,a2-pi0pi0}.
Additionally, to coalesce into a deuteron, the two $\pi^0$s should 
be emitted so as to compensate for the momentum given 
to the QF participant nucleon.
Therefore, every FA calculation yields an enhancement
in the central region of the spectrum.
The $d\sigma/dM_{\pi{d}}$ shows two peaks.
The centroid of the low-mass peak is $\sim${2.15} GeV$/c^2$ 
independently of the incident energy.
However, that of the high-mass peak decreases with a decrease
in the incident energy, and finally the two peaks are merged into a bump.
The high-mass peak reflects the appearance of the 2.15-GeV$/c^2$ 
peak in $d\sigma/dM_{\pi{d}}$ between the other pion and deuteron (reflection).
This interpretation becomes obvious by looking at 
Fig.~\ref{fig3} which shows the correlation between $W_{\gamma d}$ and $M_{\pi d}$ 
and that between two $M_{\pi d}$s.
The loci corresponding to $M_{\pi d}=2.15$ GeV$/c^2$ are clearly observed.
It should be noted that 
the incident energy limits the area in each correlation.
The $M_{\pi d}$ spectra 
of the FA calculations shown in Fig.~\ref{fig2} (central)
 also show a similar spectrum having two peaks,
which are caused by QF $\pi^0\Delta$ production.

\begin{figure}[htbp]
\begin{center}
\includegraphics[width=0.9\textwidth]{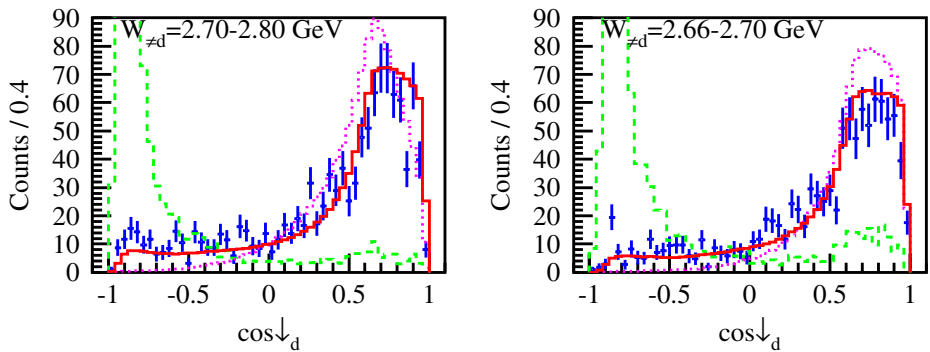}
\end{center}
\caption{Raw angular distributions of deuteron emission  (each distribution 
corresponds to the number of counts
without acceptance correction) in the $\gamma d$-CM frame
for the two highest-energy photon-tagging groups.
The dashed (green) and dotted (magenta) curves 
correspond to the FA calculation (QFC process) and semi-QF  process.
The solid curves  (red) show the pure phase-space process.
Here, the detector acceptance of each process is taken into account,
and overall normalization is made so that the total counts 
(namely the area) should be the same as that for the experimental data.
}\label{fig4}
\end{figure}

A great difference between the FA calculation based on the QFC process
and the experimental data 
appears in the angular distribution of $d\sigma/d\Omega_d$ in Fig.~\ref{fig2} (left).
The experimentally obtained $d\sigma/d\Omega_d$ shows 
a gradually-increasing behavior with decrease of  $\cos\theta_d$.
While the FA calculation provides 
a strongly backward-peaking behavior.
The main difference is observed at extremely backward angles where 
the acceptance is very low for the events.
To confirm the difference is not originated from 
the detector acceptance,
the raw angular distributions are also compared between
the FA calculation and data.
Fig.~\ref{fig4} shows the raw angular distributions (each distribution corresponds to the number of counts
without acceptance correction) of deuteron emission in 
the $\gamma d$-CM frame for the two highest-energy photon-tagging groups. 
The yield of the experimental data is relatively small
at forward angles than that of the phase-space generation 
(solid curve in red) corresponding to $d\sigma/d\Omega_d$ shown in Fig.~\ref{fig2} (left).
The difference is small between the distributions for the 
experimental data and pure phase-space generation.
The FA calculation (dashed curve in green) shows a significant enhancement 
in a wide range of backward  angles,
and does not reproduce the experimental data

The following semi-QF process may be considered: the first $\pi^0$ is emitted from the QF nucleon,
subsequently the $NN$ ($\Delta{N}$) reaction 
occurs with the spectator nucleon to generate 
${\cal D}_{12}$ with a mass of 2.15 GeV/$c^2$, followed by 
the second $\pi^0$ and deuteron emission. The kinematics of this process, however, 
creates a sideway peak in $d\sigma/d\Omega_d$ at high incident energies
as shown in Fig.~\ref{fig2} (dotted curve in magenta). 
Apparently, the semi-QF process does not reproduce the experimental distribution.

\begin{figure}[htbp]
\begin{center}
\includegraphics[width=0.9\textwidth]{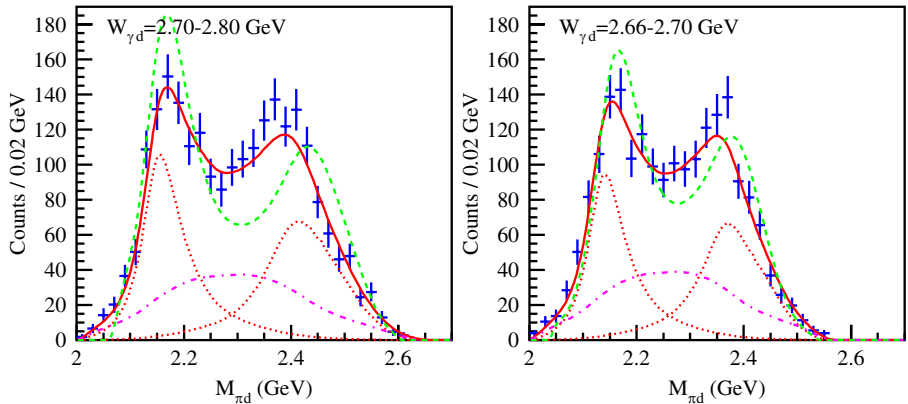}
\end{center}
\caption{
$M_{\pi{d}}$ spectra
for the two highest-energy photon-tagging groups.
The solid curves (red) show the fitted functions, 
expressed as a sum of a Breit-Wigner (BW) peak (dotted curves in red), its reflection
(dotted curves in red), and phase-space (dash-dotted curves in magenta) contributions,
 to the data.
Contributions of  a BW peak and its reflection are summed up at an amplitude level.
The dashed curves (green) correspond to the FA calculations
with arbitrary normalization.
}\label{fig5}
\end{figure}

We conclude that the 
peak at 2.15 GeV$/c^2$ in the  $M_{\pi{d}}$ spectrum is attributed to a dibaryon state, 
which can be generated in neither the QFC process nor the semi-QF process. 
This conclusion motivates the following interpretation for the resonance-like structure
in Fig.~\ref{fig1}: a dibaryon state can be formed from a deuteron and a photon; 
the generated state may be a loosely-coupled molecular state; it plays a role as a doorway to a more 
complicated dibaryon state. Of particular importance is
the fact that there is no spectator nucleon in the observed reaction.
We fit a function expressed by 
a sum of three Breit-Wigner (BW) peaks and phase-space contributions
to the data.
The function is given by
\begin{equation}
\sigma(W_{\gamma{d}}) = \sigma_{\rm{PS}}(W_{\gamma{d}})
\left\{
1+
\sum_{i=0}^2
\alpha_i L_{M_i,\Gamma_i}(W_{\gamma{d}}/c^2)
\right\},
\label{eq:3bw}
\end{equation}
where $\sigma_{\rm{PS}}(W_{\gamma{d}})$ 
denotes $\sigma$ for the phase-space contribution, 
$L_{M,\Gamma}(W_{\gamma{d}})$ represents a
BW function with the centroid of $M$ and width of $\Gamma$.
The $M_0${$=$}$2.37$ GeV/$c^2$ and $\Gamma_0${$=$}$0.07$ GeV/$c^2$ are 
fixed to the values for $d^*(2380)$~\cite{cosy1},
and the absolute values for $\sigma_{\rm{PS}}(W_{\gamma{d}})$ are 
determined to fit the phase-space component 
in each $M_{\pi d}$ spectrum (see Fig.~\ref{fig5}).
The fitted function and each contribution to it are also plotted in Fig.~\ref{fig1}.
The obtained parameters for the two peaks 
are 
$(M_1,\Gamma_1)${$=$}$(2.469{\pm}0.002, 0.120{\pm}0.003)$ GeV/$c^2$ and
$(M_2,\Gamma_2)${$=$}$(2.632{\pm}0.003, 0.132{\pm}0.005)$ GeV/$c^2$.
Isoscalar dibaryons appear not only at 2.38 GeV/$c^2$ but also at 2.47 and 2.63 GeV/$c^2$. 
It should be noted that each observed peak may be comprised of several overlapping resonances
and that the WASA-at-COSY data covering masses up to 2.56 GeV$/c^2$
do not show any signature of the 2.47-GeV$/c^2$ dibaryon.

To study  the properties of the 2.15-GeV$/c^2$ peak 
in  $d\sigma/dM_{\pi{d}}$, we have analyzed the $M_{\pi{d}}$ spectra for the two highest-energy 
photon-tagging groups as shown in Fig~\ref{fig5} in more detail. 
Here, we consider the sequential and non-sequential processes of two-$\pi^0$
emission. The contribution from the latter process
is assumed to be proportional to the phase space.
At first, the mass and width are determined by  fitting
a function, expressed as a sum of a BW-peak,
its reflection, and phase-space contributions, to the $M_{\pi{d}}$ data.
The function is given by convolution of a Gaussian with an experimental mass resolution 
of $\sigma_M=0.011$ GeV$/c^2$, and
\begin{equation}
N (m_1)  = \int_{m_2}
\left(\alpha  \left| L_{M,\Gamma}(m_1)+L_{M,\Gamma}(m_2) \right|^2+C\right) 
V_{\rm PS}(m_1, m_2) \, dm_2,
\end{equation}
where $V_{\rm{PS}}(m_1,m_2)$
expresses the phase-space contribution, 
$L_{M,\Gamma}(m) = \left(m^2-M^2+ i M\Gamma\right)^{-1}$ represents the
BW amplitude with $M$ and $\Gamma$. 
The acceptance is taken into account 
depending on  $M_{\pi\pi}$, $M_{\pi{d}}$, and $\cos\theta_d$
to estimate $V_{\rm{PS}}(m_1,m_2)$.
The parameters obtained are
$M=2.140{\pm}0.011$ GeV$/c^2$ and  $\Gamma${$=$}$0.091{\pm}0.011$ GeV$/c^2$.
The mass is slightly lower
than the sum of the $N$ and $\Delta$ masses
($\sim$2.170 GeV/$c^2$),
and  the width is narrower than that of 
$\Delta$
($\sim$0.117 GeV/$c^2$)~\cite{pdg}.
It should be noted that the FA calculation does not reproduce the $M_{\pi d}$ distribution, either. The centroid of the second peak is higher 
at higher incident energy.

\begin{figure}[htbp]
\begin{center}
\includegraphics[width=0.9\textwidth]{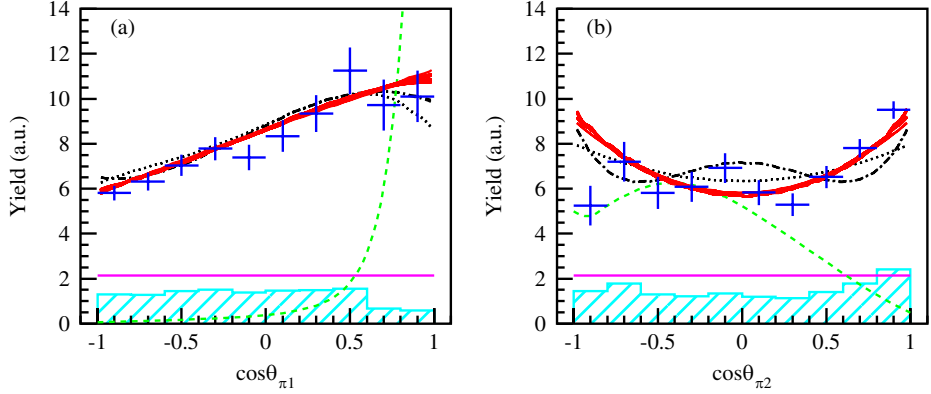}
\end{center}
\caption{
Acceptance-corrected angular distributions
for $\pi_1$ in the $\gamma{d}$-CM frame 
($z$ axis: the photon beam direction) (a),
and for $\pi_2$ in the $\pi_2{d}$ rest frame
($z$ axis: the opposite direction to $\pi_1$) (b).
Events with $M_{\pi_2{d}}${$=$}$2.05$--2.25 GeV$/c^2$ and $W_{\gamma{d}}${$=$}2.66--2.80 GeV
 are selected.
The lower hatched histograms (cyan) show 
the corresponding systematic errors.
The dashed curves (green) show the corresponding distributions
in the FA calculations.
The angular distributions are plotted with a shaded band (red)
for $J_2^\pi${$=$}$1^+$, $2^+$, and $3^-$,
and with dotted and dash-dotted curves (black)
for $J_2^\pi${$=$}$1^-$ and $2^-$, respectively.
The solid horizontal lines (magenta) show the 
phase-space contributions.
}\label{fig6}
\end{figure}

Information on the spin-parity of the dibaryon states has been 
deduced from angular distributions of $\pi^0$s obtained for 
the events from $M_{\pi{d}}${$=$}$2.05$--2.25 GeV$/c^2$ in Fig.~\ref{fig5}.
We define $\pi_1$ and $\pi_2$ as follows:
$\pi_2$ is one of the two pions giving $2.05\le M_{\pi d}<2.25$ GeV$/c^2$, 
and $\pi_1$ is the other pion.
Here, we assume $\pi_1$ is emitted first, leaving the $\pi_2d$ system
with $2.05\le M_{\pi d}<2.25$ GeV$/c^2$, and $\pi_2$ is emitted subsequently.
We combine the data of the two highest-energy 
groups to analyze angular distributions.
Fig.~\ref{fig6}(a) shows the deduced $\pi_1$ angular distribution in
 the $\gamma{d}$-CM frame with the $z$ axis taken along 
the incident photon direction. 
The 
experimental distribution is mostly expressed by a sum of two terms, constant and proportional to $\cos\theta$. This $\cos\theta$ dependence is naturally understood as a result of interference between $\pi_1$-emission amplitudes with different parities. 
The $\pi_2$
angular distribution in the rest frame of the $\pi_2 d$ system is shown in Fig.~\ref{fig6}(b), where the $z$ axis is defined to be opposite to the $\pi_1$-emission direction. 
Unlike Fig.~\ref{fig6}(a), the distribution shows almost $90^\circ$ symmetry. This implies that the $2.15$-GeV/$c^2$ resonance is 
made of a single $J^\pi$ state or mixed
states with the same parity.
The FA calculation completely fails to reproduce both the
angular distributions, due to the difference in the underlying reaction 
mechanism as discussed previously. 
A sharp peak at $0^\circ$ in Fig.~\ref{fig6}(a) is the reflection of 
the backward peak in $d\sigma/d\Omega_d$. In Fig.~\ref{fig6}(b), the distribution 
takes an upward-convex shape being opposite to the experiment.

We calculate the $\pi_1$ and $\pi_2$
angular distributions for
the reaction sequence $\gamma{d}${$\to$}${R_1}${$\to$}$\pi_1{R_2}${$\to$}$\pi_1\pi_2{d}$
using the density matrix (statistical tensor) formalism~\cite{rose},
where spins of $R_1$ and $R_2$ are denoted by $J_1$ and $J_2$, respectively.
The formalism incorporates the exchange symmetry between $\pi_1$ and $\pi_2$.
A possible contribution from non-sequential $\pi^0\pi^0$ production is 
assumed to be proportional to 
the phase-space contribution,
of which the fraction is determined from the $M_{\pi{d}}$ spectrum in 
Fig.~\ref{fig5} for each photon-tagging group.
Since the $\pi_2$ angular distribution  is almost symmetric with respect to $\cos\theta_{\pi_2}${$=$}$0$, the interference effect may be small. Thus,
the contributions from the sequential and non-sequential processes are summed 
up incoherently.
A set of the amplitudes
of sequential processes $A_{\Lambda\Lambda}$ is determined 
for all the $\Lambda${$=$}$(L_0,J_1,L_1,J_2,L_2)$ combinations
to reproduce the measured $\pi_1$ and $\pi_2$ angular distributions
simultaneously (20 data points),
where $L_0$, $L_1$, and $L_2 \le 2$
denote angular momenta carried by the incident photon,
$\pi_1$ emission, and $\pi_2$ emission,
respectively.
An amplitude for a mixed state is given
by $A_{\Lambda\Lambda'}${$=$}$\left(A_{\Lambda\Lambda}A_{\Lambda'\Lambda'}\right)^{1/2}$.
The $S$-wave $NN^*$ molecular states
are assumed to play a role as a doorway to $R_1$.
Here, the considered $N^*$s are $D_{15}(1675)$, $F_{15}(1680)$, 
 and $P_{13}(1720)$, 
which give dominant contributions in the $\gamma{N}${$\to$}$\pi^0\pi^0{N}$ reaction 
in the relevant energy region~\cite{a2-pi0pi0}.
Hence, 
$J_1^\pi$s
under consideration are $1^+$, $2^\pm$, and $3^\pm$.
Additionally, $R_2$ is assumed to be a single resonance,
namely $J_2${$=$}$J_2'$ and $L_2${$=$}$L_2'$.

In Fig.~\ref{fig6}, also shown are the angular distributions calculated for 
$J_2^\pi${$=$}$1^\pm$, $2^\pm$, and $3^+$. The $J_2^\pi${$=$}$0^\pm$ assignments 
are already excluded because of an isotropic distribution for $\pi_2$ emission. 
The assignments of $J_2^\pi${$=$}$1^+$, $2^+$, and $3^-$ show almost the same 
quality to reproduce the angular distributions ($29<\chi^2<32$).
We reject the $J_2^\pi${$=$}$1^-$ and $2^-$ assignments,
giving worse distributions 
($\chi^2${$=$}$47$ and 55), respectively, with 
a confidence level of higher than 99.7\% ($3\sigma$), and leave 
the possibility of $J_2^\pi${$=$}$1^+$, $2^+$ and $3^-$. 
Regarding $J_1^\pi$, major components are 
$1^+$ (${\sim}$70\%) and
$2^-$ (${\sim}$20\%) for the case of $J_2^\pi${$=$}$1^+$ and $2^+$, 
while they are distributed widely to $J_1^\pi${$=$}$1^+$, $2^\pm$, and $3^+$ for 
$J_2^\pi${$=$}$3^-$. The $J_2^\pi${$=$}$2^+$ assignment not only coincides with the energy dependence of the 
$^3\!P_2$-$\pi{d}$ amplitude,
but also supports the existence of the predicted ${\cal{D}}_{12}$ state at 2.15 GeV$/c^2$.
The $J_2^\pi${$=$}$3^-$ assignment is consistent with
the energy dependence of the $^3\!D_3$-$\pi{d}$ amplitude (about a half
strength of $^3\!P_2$).
There is no experimental sign for 
a $1^+$ state, although 
two isovector $0^-$ and $2^-$ states at 2.2 GeV$/c^2$ 
have been reported recently~\cite{anke}.

In summary,
the total and differential cross sections have been measured 
for the $\gamma{d}${$\to$}$\pi^0\pi^0{d}$ reaction
at $E_\gamma${$=$}$0.75$--1.15 GeV.
The total cross section as a function of $W_{\gamma{d}}$
shows resonance-like behavior, which peaks at approximately 2.47 and 2.63 GeV.
The theoretical calculation 
based on the QFC mechanism reproduces its behavior well.
However, the experimental angular distributions 
of deuteron emission can never be understood in the QFC mechanism. 
A possible scenario is that one nucleon in the deuteron is excited 
by photoabsorption but is still interacting with the other nucleon
before emitting two $\pi^0$s,
forming a dibaryon resonance.
In $\pi^0d$ invariant-mass distributions
corresponding to the state after emitting the first $\pi^0$,
a clear peak is observed 
at $2.14{\pm}0.01$ GeV$/c^2$ 
with a width of $0.09{\pm}0.01$ GeV$/c^2$.
The angular distributions for the two $\pi^0$s 
limit  $J^\pi$ of the state to $1^+$, $2^+$, or $3^-$.
The $2^+$ assignment is consistent with
the theoretically predicted ${\cal{D}}_{12}$ state,
and with the resonance structure of 
the $^3\!P_2$-$\pi{d}$ amplitude.
The present work shows 
strong evidence for the existence of the 2.15-GeV$/c^2$ isovector 
dibaryon in the $\pi^0d$ channel,
and of the 2.47- and 2.63-GeV$/c^2$ isoscalar dibaryons 
in the $\pi^0\pi^0d$ channel.
These findings would give a base to explore dibaryon states 
lying at higher masses.

The authors express their gratitude to the ELPH accelerator staff for stable operation 
of the accelerators in the FOREST experiments.
They acknowledge Mr.\ Kazue~Matsuda, Mr.~Ken'ichi~Nanbu, and Mr.~Ikuro~Nagasawa for their technical assistance in the FOREST experiments.
They also thank Prof.\ Alexander~I.~Fix for the theoretical calculations
of the cross sections and fruitful discussion. 
This work was supported in part by the Ministry of Education, Culture, Sports, Science and Technology, Japan
through Grants-in-Aid for Scientific Research (B) No.\ 17340063, 
for Specially Promoted Research No.\ 19002003,
for Scientific Research (A) No.\ 24244022,
for Scientific Research (C) No.\ 26400287,
and for Scientific Research (A) No.\ 16H02188.


\end{document}